\documentclass[%
 aip,
 amsmath,amssymb,
 reprint,%
]{revtex4-1}

\usepackage{graphicx}
\usepackage{dcolumn}
\usepackage{bm}

\draft 

\begin{document}


\title{Coherent high frequency axial oscillations in a partially magnetized direct current magnetron discharge} 

\author{R. C. Przybocki}
\email{ryancp@stanford.edu}
\author{M. A. Cappelli}

\affiliation{Stanford Plasma Physics Laboratory, Department of Mechanical Engineering, Stanford University, Stanford, CA, 94305, USA}

\date{\today}

\begin{abstract}

High frequency oscillations are observed in a neon plasma of a direct current magnetron discharge.  At low discharge currents, we see highly coherent 60 MHz fluctuations.  Above a distinct current threshold, secondary $5 - 10$ MHz fluctuations emerge in addition to turbulent fluctuations in the $60 – 100$ MHz range.  The oscillations in the total discharge current suggest axial wave propagation.  A lower-hybrid wave theory is invoked to model the high frequency oscillations.  We attribute the low frequency modes to a turbulence-driven inverse cascade process, as suggested by recent simulations.
\end{abstract}

\maketitle 

Low pressure plasma discharges frequently employ perpendicular electric (E) and magnetic (B) fields to confine electrons along the E$\times$B drift direction. E$\times$B devices such as Hall thrusters, Penning discharges, high-power impulse magnetron sputtering devices, and direct current (DC) magnetrons, consequently, tend to exhibit fluctuations in plasma properties \cite{mcdonald, rodriguez2019, held, ito2015, panjan2017plasma, choueiri2001}, the origins of which are not well understood and the subject of debate \cite{boeuf20, hara22}.  Previous studies on magnetrons \cite{ito2009, ito2015, marcovati2020, marcovati2023, przybocki2023, hecimovic, boeuf2023, panjan2015, revel2016} have observed long-wavelength, azimuthally-propagating disturbances at frequencies of $0.1 - 3$ MHz, attributed to gradients in density and magnetic field.  In this letter, we report on disturbances with frequencies of $60 - 100$ MHz in a neon DC magnetron discharge. To our knowledge, these disturbances, which are quite coherent in some cases, have not been previously reported.  We attribute them to lower-hybrid waves propagating nearly axially, which are destabilized by particle drifts, collisions, and gradients in density and magnetic field.  We also find that a transition occurs at higher discharge currents that leads to the emergence of microturbulence and the appearance of coherent low frequency modes.  We speculate that these two features are connected.

\begin{figure}
\includegraphics{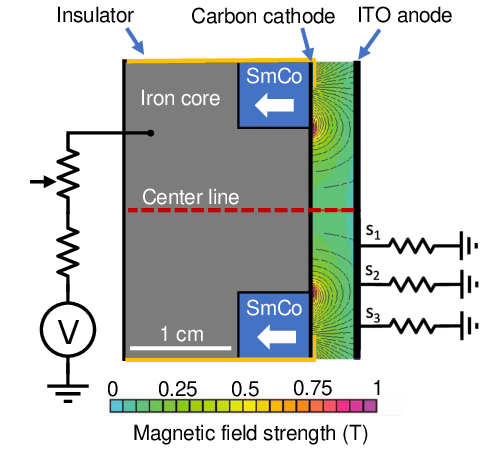}
\caption{\label{fig1} Experimental setup, showing electrical connections to the anode segments $s_1$, $s_2$, and $s_3$.  The magnetic field topology is simulated using the Finite Element Method Magnetics (FEMM) software \cite{baltzis2008}.  Note that the gap shown to illustrate the topology is $5$ mm, which is larger than the 2 mm value used in the experiments.}
\end{figure}

A schematic of the discharge is shown in Fig. \ref{fig1}.  For a detailed description, we direct the reader to previous studies \cite{marcovati2023, przybocki2023}. We use the same apparatus, which operates at low pressure within a vacuum chamber. A 35 mm diameter samarium-cobalt (SmCo) ring magnet on an iron core produces an azimuthally symmetric magnetic field with topology between the discharge gap depicted in the figure. The highest plasma density is expected where the magnetic field is radial and electrons are confined by the E$\times$B drift.  A graphite sheet serves as the cathode to minimize sputtering, and the anode consists of a glass plate coated with indium tin oxide (ITO).  The cathode-anode gap is fixed at 2 mm.  The ITO anode is divided into two electrically isolated 10-degree segments and one isolated 340-degree segment to characterize the azimuthal propagation of plasma disturbances. 

The ambient neon gas pressure is maintained at 27 Pa.  Above a breakdown voltage of $\sim 360$ V, plasma forms in a 19 mm diameter ring confined near the magnet where there is a strong radial field component. The current measured at the anode shows strong oscillations, indicating the presence of oscillating plasma structures.

\begin{figure*}
    \includegraphics{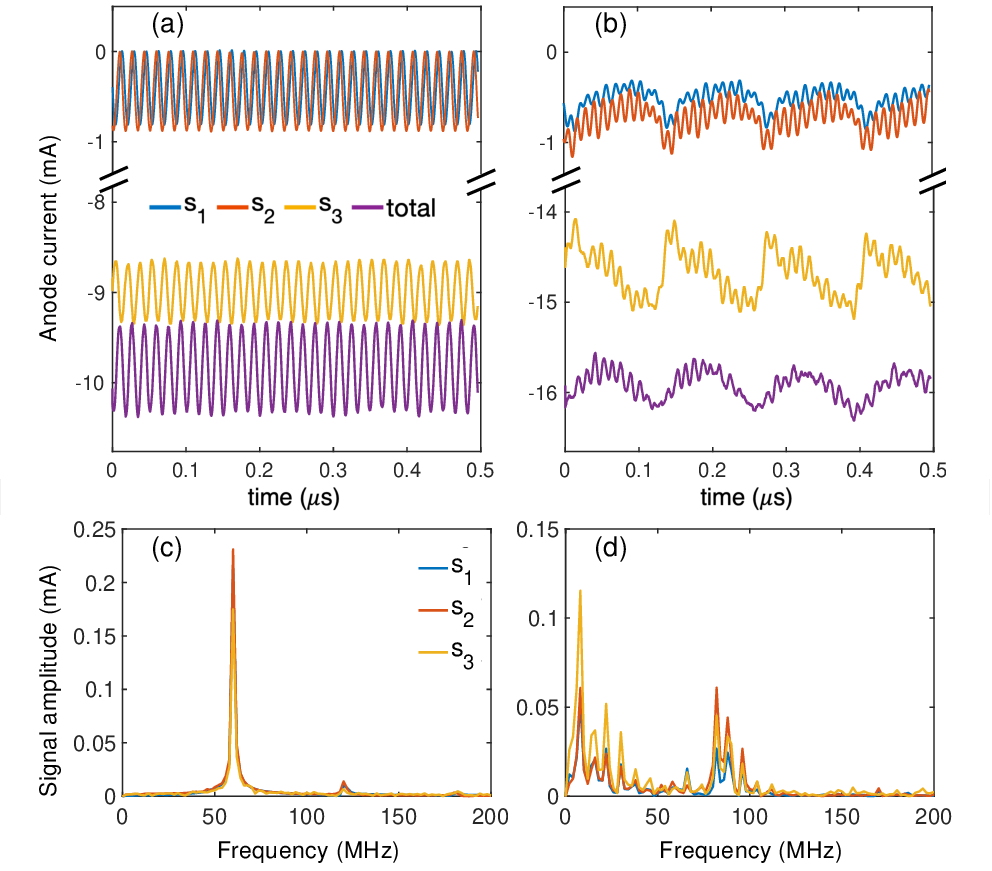}
    \caption{\label{current_fft}  Current traces from anode segments and the corresponding Fast Fourier Transforms (FFT's) of the signals.  $s_1$ and $s_2$ are the 10\textdegree \ segments while $s_3$ is the 340\textdegree \ segment measuring the remainder of the current.  The total current is the sum of the segments, shown by the purple trace.  The discharge conditions are (a) $V = 366.2$ V, $\overline{I} = 9.85$ mA; (b) $V = 396.0$ V, $\overline{I} = 15.92$ mA.  The respective FFT's of the signals are plotted in (c) and (d) to illustrate the frequency content of the data. The signal in (a) is dominated by a frequency component $\approx 60$ MHz, with harmonics visible at $\approx$ 120 and 180 MHz.  The signal in (b) contains two distinct frequency peaks at $\approx 8$ MHz and $\approx$ 80 MHz.  Additional smaller amplitude peaks are seen with a spacing of $6 - 8$ MHz.}  
\end{figure*}

\begin{figure}
    \includegraphics{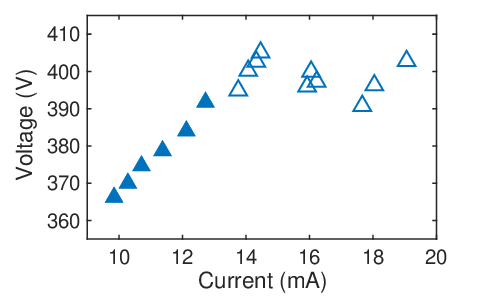}
    \caption{\label{iv_fig} Voltage as a function of current for the discharge.  The single and dual frequency operating regimes are denoted by the filled and open data points, respectively.}
\end{figure}

We identify two operating regimes for this magnetron discharge.  The first is characterized by coherent oscillations in the measured current at a single fundamental frequency of about 60 MHz.  We refer to this as the "single frequency" regime.  An example is shown by the current traces in Fig. \ref{current_fft}(a), with the corresponding Fast Fourier Transform (FFT) in Fig. \ref{current_fft}(c).  The discharge voltage and mean total current in this example are $V = 366.2$ V and $\overline{I} = 9.85$ mA.  The total current (given by the sum of the three segments) oscillates at the same frequency as the segments, suggesting a plasma disturbance propagating along the discharge axis.  The oscillations in the total current were also observed with a Pearson model 2877 current monitor on the cathode side to rule out the possibility of signal delay effects.  The oscillations in the two 10\textdegree \ segments ($s_{1}$ and $s_{2}$) are out of phase, indicating an azimuthal component to the wave propagation.

The second operating regime is distinguished from the first by significant lower frequency ($5 - 10$ MHz) fluctuations along with oscillations at $60 - 100$ MHz.  We refer to this as the "dual frequency" regime.  Both frequency ranges are visible in the current traces of Fig. \ref{current_fft}(b), measured with discharge conditions $V = 396.0$ V and $\overline{I} = 15.92$ mA.  The FFT contains the largest components at about 8 MHz and 80 MHz.  Other lower amplitude components are visible with successive peaks spaced $6 - 8$ MHz apart.  

The discharge voltage and current conditions across both operating regimes are shown in Fig. \ref{iv_fig}.  The single frequency regime occurs for current $\overline{I} < 13$ mA and exhibits a positive voltage coefficient of resistance with the voltage increasing nearly linearly with current.  The dual frequency regime appears for $\overline{I} > 13$ mA, with a voltage that is relatively insensitive to current, perhaps decreasing slightly with increasing current.  We hypothesize that this transition in the voltage coefficient of resistance results from the onset of microturbulence at higher power, which breaks the plasma confinement and causes a gradual increase in conductivity and cross-field electron mobility.  Evidence of turbulence has previously been reported in experiments with argon \cite{marcovati2023} and nitrogen \cite{przybocki2023} and in E$\times$B plasma simulations \cite{smolyakov_prl, tyushev}.  

\begin{figure}
    \includegraphics[]{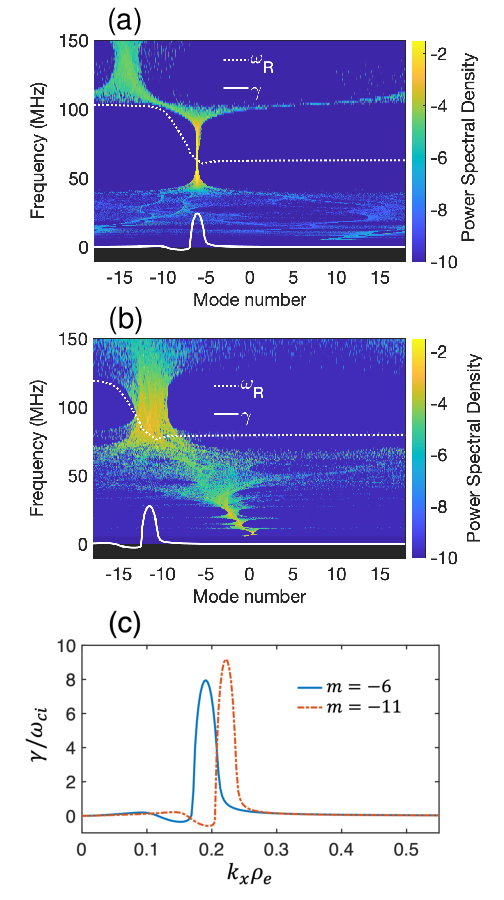}
    \caption{\label{dispersion_fig} Plots (a) and (b): power spectral density maps of the signals shown in Figs. \ref{current_fft}(a) and (b).  The data is normalized to the total power in the spectrum and plotted on a logarithmic scale.  The discharge conditions are (a) $V = 366.2$ V, $\overline{I} = 9.85$ mA; (b) $V = 396.0$ V, $\overline{I} = 15.92$ mA.  Plot (a) shows the single frequency condition with oscillations at $m = -6$.  Plot (b) contains two high-density regions: one at low frequencies with mode numbers near $m = 0$ and another centered at $m = -11$ at frequencies of $75 - 100$ MHz.  Also shown is the $\gamma > 0$ solution branch to Eq. \ref{dispersion}.  The frequency $\omega_{R}$ is plotted by the dotted line and the growth rate $\gamma$ by the solid line.  Plot (c): solution for $\gamma$ as a function of $k_x$, used to determine $k_x$ in the dispersion model.  The input parameters are the same for the two curves except for the choice of $m$ and $v_{e0,x}$, which reflect the different conditions for the single vs. dual frequency regimes.}
\end{figure}

Following the approach described in prior work\cite{marcovati2023, przybocki2023}, we identify the fluctuations' azimuthal components through a wavelet decomposition analysis on the measured segment currents.  The azimuthal wavenumber at each frequency is determined from the phase shift between the complex signals from the two 10\textdegree \ segments.  The signal is mapped to a power spectral density in frequency-wavenumber space, which shows the frequencies and wavenumbers that carry the most energy.  The power spectra for the signals of Fig. \ref{current_fft} are shown in Fig. \ref{dispersion_fig}, normalized to the total power in the spectrum.  We denote the azimuthal direction as $y$ (counterclockwise as viewed from anode to cathode) and introduce the non-dimensional wavenumber (mode number) $m = k_{y} R_{0}$, where $R_0$ is the plasma radius.  A positive mode number designates a wavevector component in the $ - \textbf{E} \times \textbf{B}$ direction, and vice versa.

The single frequency fluctuations occur with frequency $f = 60-65$ MHz at $m = -6$, corresponding to a component wavelength $\lambda_{y} \approx 1$ cm.  This is visible in the power spectral density plot of Fig. \ref{dispersion_fig}(a) as a narrow region at $m = -6$.  For the dual frequency case, the low frequency oscillations have the smallest mode numbers ($m \approx -3$ to 0), and the high frequency components are in the range $m = -6$ to $-16$, centered at $m = -11$.  The high frequency oscillations in Fig. \ref{dispersion_fig}(b) are more diffuse in wavenumber space, occupying a width of several mode numbers.  We suggest that this loss of coherence is connected to the turbulent behavior invoked to explain the voltage-current characteristics.

We interpret the experiments through a fluid model for axial-azimuthal lower-hybrid modes in a partially magnetized plasma described by Smolyakov et al. \cite{smolyakov2017}   Lower-hybrid waves have been studied in laboratory and space plasmas, where an instability is driven by cross-field currents and gradients in density and magnetic field \cite{mikellides, krall, norgren, huba1978, davidson1976, davidson1975}.  These waves propagate perpendicular to the magnetic field ($\mathbf{k} \cdot \mathbf{B} = 0$) in the frequency range $\omega_{ci}< \omega < \omega_{ce}$. 

The experiment is modeled as follows: the electric field is axial, $\textbf{E} = E_{0} \hat{x}$, the magnetic field is radial, $\textbf{B} = B_{0} \hat{z}$, and $y$ is the azimuthal direction.  We include equilibrium electron and ion drift velocities $\textbf{v}_{e,i0}$.  In the obstructed discharge of this experiment, the equilibrium plasma density $n_0$ exhibits strong axial gradients that are modeled with the characteristic length $L_n = n_0 (dn_{0}/dx)^{-1}$.  Similarly, the radial magnetic field strength varies axially and is modeled with the characteristic length $L_B = B_0 (dB_{0}/dx)^{-1}$.  The perturbations in plasma properties $\psi$ are modeled as $\tilde{\psi} \sim \textrm{exp}(i \textbf{k} \cdot \textbf{x} - i \omega t)$ with wavevector $\textbf{k} = (k_x, k_y, 0)$ and complex frequency $\omega$.  Without the assumption of quasineutral perturbations, the dispersion relation is\cite{smolyakov2017}:

\begin{multline}
    \label{dispersion}
    0 = k^{2}\lambda_{D}^{2} - \frac{k^{2}c_{s}^2}{(\omega - \omega_{i0})^{2}} \\
    + \frac{\omega_{*} - \omega_{D} + k^{2} \rho_{e}^{2}(\omega - \omega_{e0} + i \nu_e)}{\omega - \omega_{e0} - \omega_{D} + k^{2} \rho_{e}^{2}(\omega - \omega_{e0} + i \nu_e)}
\end{multline}
where $k = (k_{x}^{2} + k_{y}^{2})^{1/2}$, $\lambda_{D} = (\epsilon_0 k_B T_e / n_0 e^{2})^{1/2}$, $c_{s} = (k_{B} T_{e} / m_{i})^{1/2}$, $\omega_{e,i0} = \textbf{k} \cdot \textbf{v}_{e,i0}$, $\nu_{e}$ is the electron-neutral collision frequency, $\omega_{*} = -k_{y} k_{B}T_{e}/eB_{0}L_{n}$, $\omega_{D} = -2 k_{y} k_{B}T_{e}/eB_{0}L_{B}$, and $\rho_{e} = \sqrt{m_e k_B T_e}/eB_{0}$.

We hypothesize that these waves originate due to excitation near the cathode, and the electrostatic field disturbances and associated current fluctuations convey back towards the anode where they are detected in our experiment.  It is not uncommon for instabilities to be triggered in regions far from where they are detected in a saturated state \cite{hara,fernandez2008}.  The plasma parameters are estimated from simulations \cite{boeuf20} and similar experiments \cite{mitic2021}: $n_0 = 2 \times 10^{17}$ m$^{-3}$, $\nu_{e} = 5 \times 10^{7}$ s$^{-1}$, and $T_{e} = 1$ eV.  The gradient length scales are $L_{n} = -0.2$ mm and $L_{B} = 0.5$ mm (the minus sign indicates plasma density decreasing near the cathode).  The electric and magnetic fields near the cathode are estimated to be $4 \times 10^{5}$ V/m and 0.65 T, respectively, giving an E$\times$B electron drift velocity $v_{e0,y} = -6.15 \times 10^{5}$ m/s.  With such strong axial gradients, when the wavelengths are comparable to the gradient length scales, the theory is limited in its application, and a computational simulation may be needed to accurately capture dynamics in the nonuniform near-cathode region. Nevertheless, we believe that the analysis presented here is valuable for interpreting the data.

We surmise that the single frequency regime occurs without significant microturbulence, so electrons are strongly confined by E$\times$B drift such that $\textbf{k} \cdot \textbf{v}_{e0} = k_y v_{e0,y}$.  The dual frequency condition appears to coincide with a change in plasma resistance characteristic of increased electron mobility and microturbulence.  For this case, we include an axial electron velocity component $v_{e0,x} = -4$ km/s, roughly 10 times the value expected from classical scattering.  The cathode-directed axial ion velocity near the sheath edge is estimated to be 10 km/s, consistent with an assumed 10 V potential drop from the anode to the sheath edge.  The Larmor radius for neon ions at this speed is 3 mm, which is larger than the discharge gap length.  We thus consider the ions to be weakly magnetized and not confined by the E$\times$B drift, so the ion velocity $\textbf{v}_{i0}$ is purely axial.  

The dispersion relation has solutions $\omega = \omega_R + i \gamma$, where $\omega_R$ is the frequency and $\gamma$ is the growth rate in the linear phase of instability.  We focus on solutions with $\gamma > 0$, which lead to growth in the perturbations.  Under the conditions described, one solution branch of Eq. \ref{dispersion} has a positive growth rate.  Since the experiment does not measure $k_x$ directly, we use the measured $m$ ($= k_y R_0$) to inform the choice of $k_x$ in the model.  Fig. \ref{dispersion_fig}(c) plots the growth rate $\gamma$ as a function of the non-dimensional axial wavenumber $k_x \rho_e$ for the two operating regimes.  The peak growth rates occur at $k_x \rho_e = 0.19$ and $0.22$ rad/m, respectively for the single and dual frequency conditions, corresponding to wavelengths $\lambda_{x,1} = 120 \ \mu m$ and $\lambda_{x,2} = 100 \ \mu m$, given an electron gyroradius $\rho_{e} = 3.7 \ \mu m$.  Since $k_x \gg k_y$, this implies that the wave propagation is predominantly axial.

The solutions to the dispersion relation are plotted as a function of $m$ in Figs. \ref{dispersion_fig}(a) and (b).  At the mode numbers of the high frequency oscillations, $\omega_R$ is 60 and 80 MHz, respectively, in agreement with the experimental results.

The origin of the lower frequency features remains uncertain.  We speculate that their appearance is tied to the onset of microturbulence at higher current, which enhances axial cross-field electron transport and reduces confinement.  This conjecture is based on simulations by Koshkarov et al. \cite{smolyakov_prl}, who observed the formation of azimuthally elongated zonal flow structures attributed to an inverse cascade from shorter wavelength instabilities.  The growth of these structures is driven by nonlinear anomalous electron current in a turbulent state.  Simulations of a Penning discharge by Tyushev et al. \cite{tyushev} also observed an inverse energy cascade to a lower frequency, longer wavelength regime at higher power, characterized by increased turbulent transport.  Turbulence in the magnetron discharge might drive the growth of lower frequency structures in a similar manner.  However, it is also possible that the onset of lower frequency structures triggers the turbulence through interactions with the high frequency modes.  High fidelity simulations may provide insight into this question of cause and effect.

To summarize, we have reported on a magnetron discharge that exhibits fluctuations of $60 - 100$ MHz in a neon plasma.  The discharge voltage characteristic transitions from a positive resistivity regime to one that is insensitive to current at a current threshold coinciding with the appearance of additional lower frequency structures.  The high frequency plasma fluctuations are attributed to axial lower-hybrid waves with a small azimuthal component.  A fluid description of lower-hybrid instabilities models the lower current regime dispersion assuming plausible plasma properties near the cathode.  Applying the same model to the higher current regime captures the shift of the disturbances to higher frequencies and wavenumbers under the assumption that microturbulence results in an axial electron drift.  We suggest that numerical computations could provide a better understanding of the growth of low frequency disturbances.  Probe measurements may also offer an additional diagnostic to describe the dynamics of the low and high frequency modes seen at high discharge currents.
\begin{acknowledgments}
The authors would like to acknowledge the assistance of A. Marcovati who constructed the experimental facility.  We would also like to thank K. Hara for his insightful discussions and advice.  Partial support for this research was provided by the Department of Energy, through Grant No. DE-SC0022084 with Dr. Nirmol Podder as Program Officer. 
\end{acknowledgments}

\providecommand{\noopsort}[1]{}\providecommand{\singleletter}[1]{#1}%


\begin{thebibliography}{15}%
\makeatletter
\providecommand \@ifxundefined [1]{%
 \@ifx{#1\undefined}
}%
\providecommand \@ifnum [1]{%
 \ifnum #1\expandafter \@firstoftwo
 \else \expandafter \@secondoftwo
 \fi
}%
\providecommand \@ifx [1]{%
 \ifx #1\expandafter \@firstoftwo
 \else \expandafter \@secondoftwo
 \fi
}%
\providecommand \natexlab [1]{#1}%
\providecommand \enquote  [1]{``#1''}%
\providecommand \bibnamefont  [1]{#1}%
\providecommand \bibfnamefont [1]{#1}%
\providecommand \citenamefont [1]{#1}%
\providecommand \href@noop [0]{\@secondoftwo}%
\providecommand \href [0]{\begingroup \@sanitize@url \@href}%
\providecommand \@href[1]{\@@startlink{#1}\@@href}%
\providecommand \@@href[1]{\endgroup#1\@@endlink}%
\providecommand \@sanitize@url [0]{\catcode `\\12\catcode `\$12\catcode
  `\&12\catcode `\#12\catcode `\^12\catcode `\_12\catcode `\%12\relax}%
\providecommand \@@startlink[1]{}%
\providecommand \@@endlink[0]{}%
\providecommand \url  [0]{\begingroup\@sanitize@url \@url }%
\providecommand \@url [1]{\endgroup\@href {#1}{\urlprefix }}%
\providecommand \urlprefix  [0]{URL }%
\providecommand \Eprint [0]{\href }%
\providecommand \doibase [0]{http://dx.doi.org/}%
\providecommand \selectlanguage [0]{\@gobble}%
\providecommand \bibinfo  [0]{\@secondoftwo}%
\providecommand \bibfield  [0]{\@secondoftwo}%
\providecommand \translation [1]{[#1]}%
\providecommand \BibitemOpen [0]{}%
\providecommand \bibitemStop [0]{}%
\providecommand \bibitemNoStop [0]{.\EOS\space}%
\providecommand \EOS [0]{\spacefactor3000\relax}%
\providecommand \BibitemShut  [1]{\csname bibitem#1\endcsname}%
\let\auto@bib@innerbib\@empty
\bibitem [{\citenamefont {McDonald}\ and\ \citenamefont
  {Gallimore}(2011)}]{mcdonald}%
  \BibitemOpen
  \bibfield  {author} {\bibinfo {author} {\bibfnamefont {M.~S.}\ \bibnamefont
  {McDonald}}\ and\ \bibinfo {author} {\bibfnamefont {A.~D.}\ \bibnamefont
  {Gallimore}},\ }\href@noop {} {\bibfield  {journal} {\bibinfo  {journal}
  {IEEE Trans. Plasma Sci.}\ }\textbf {\bibinfo {volume} {39}},\ \bibinfo
  {pages} {2952} (\bibinfo {year} {2011})}\BibitemShut {NoStop}%
\bibitem [{\citenamefont {Rodriguez}\ \emph {et~al.}(2019)\citenamefont
  {Rodriguez}, \citenamefont {Skoutnev}, \citenamefont {Raitses}, \citenamefont
  {Powis}, \citenamefont {Kaganovich},\ and\ \citenamefont
  {Smolyakov}}]{rodriguez2019}%
  \BibitemOpen
  \bibfield  {author} {\bibinfo {author} {\bibfnamefont {E.}~\bibnamefont
  {Rodriguez}}, \bibinfo {author} {\bibfnamefont {V.}~\bibnamefont {Skoutnev}},
  \bibinfo {author} {\bibfnamefont {Y.}~\bibnamefont {Raitses}}, \bibinfo
  {author} {\bibfnamefont {A.}~\bibnamefont {Powis}}, \bibinfo {author}
  {\bibfnamefont {I.}~\bibnamefont {Kaganovich}}, \ and\ \bibinfo {author}
  {\bibfnamefont {A.}~\bibnamefont {Smolyakov}},\ }\href@noop {} {\bibfield
  {journal} {\bibinfo  {journal} {Phys. Plasmas}\ }\textbf {\bibinfo {volume}
  {26}},\ \bibinfo {pages} {053503} (\bibinfo {year} {2019})}\BibitemShut
  {NoStop}%
\bibitem [{\citenamefont {Held}, \citenamefont {George},\ and\ \citenamefont
  {von Keudell}(2022)}]{held}%
  \BibitemOpen
  \bibfield  {author} {\bibinfo {author} {\bibfnamefont {J.}~\bibnamefont
  {Held}}, \bibinfo {author} {\bibfnamefont {M.}~\bibnamefont {George}}, \ and\
  \bibinfo {author} {\bibfnamefont {A.}~\bibnamefont {von Keudell}},\
  }\href@noop {} {\bibfield  {journal} {\bibinfo  {journal} {Plasma Sources
  Sci. Technol.}\ }\textbf {\bibinfo {volume} {31}},\ \bibinfo {pages} {085013}
  (\bibinfo {year} {2022})}\BibitemShut {NoStop}%
\bibitem [{\citenamefont {Ito}, \citenamefont {Young},\ and\ \citenamefont
  {Cappelli}(2015)}]{ito2015}%
  \BibitemOpen
  \bibfield  {author} {\bibinfo {author} {\bibfnamefont {T.}~\bibnamefont
  {Ito}}, \bibinfo {author} {\bibfnamefont {C.~V.}\ \bibnamefont {Young}}, \
  and\ \bibinfo {author} {\bibfnamefont {M.~A.}\ \bibnamefont {Cappelli}},\
  }\href@noop {} {\bibfield  {journal} {\bibinfo  {journal} {Appl. Phys.
  Lett.}\ }\textbf {\bibinfo {volume} {106}},\ \bibinfo {pages} {254104}
  (\bibinfo {year} {2015})}\BibitemShut {NoStop}%
\bibitem [{\citenamefont {Panjan}\ and\ \citenamefont
  {Anders}(2017)}]{panjan2017plasma}%
  \BibitemOpen
  \bibfield  {author} {\bibinfo {author} {\bibfnamefont {M.}~\bibnamefont
  {Panjan}}\ and\ \bibinfo {author} {\bibfnamefont {A.}~\bibnamefont
  {Anders}},\ }\href@noop {} {\bibfield  {journal} {\bibinfo  {journal} {J.
  Appl. Phys.}\ }\textbf {\bibinfo {volume} {121}},\ \bibinfo {pages} {063302}
  (\bibinfo {year} {2017})}\BibitemShut {NoStop}%
\bibitem [{\citenamefont {Choueiri}(2001)}]{choueiri2001}%
  \BibitemOpen
  \bibfield  {author} {\bibinfo {author} {\bibfnamefont {E.}~\bibnamefont
  {Choueiri}},\ }\href@noop {} {\bibfield  {journal} {\bibinfo  {journal}
  {Phys. Plasmas}\ }\textbf {\bibinfo {volume} {8}},\ \bibinfo {pages} {1411}
  (\bibinfo {year} {2001})}\BibitemShut {NoStop}%
\bibitem [{\citenamefont {Boeuf}\ and\ \citenamefont
  {Takahashi}(2020)}]{boeuf20}%
  \BibitemOpen
  \bibfield  {author} {\bibinfo {author} {\bibfnamefont {J.~P.}\ \bibnamefont
  {Boeuf}}\ and\ \bibinfo {author} {\bibfnamefont {M.}~\bibnamefont
  {Takahashi}},\ }\href@noop {} {\bibfield  {journal} {\bibinfo  {journal}
  {Phys. Rev. Lett.}\ }\textbf {\bibinfo {volume} {124}},\ \bibinfo {pages}
  {185005} (\bibinfo {year} {2020})}\BibitemShut {NoStop}%
\bibitem [{\citenamefont {Hara},\citenamefont{Mansour},\ and\ \citenamefont
  {Tsikata}(2022)}]{hara22}%
  \BibitemOpen
  \bibfield  {author} {\bibinfo {author} {\bibfnamefont {K.}\ \bibnamefont
  {Hara}}, \bibinfo {author} {\bibfnamefont {A.}~\bibnamefont {Mansour}},\ and\ \bibinfo {author} {\bibfnamefont {S.}~\bibnamefont
  {Tsikata}},\ }\href@noop {} {\bibfield  {journal} {\bibinfo  {journal}
  {J. Plasma Phys.}\ }\textbf {\bibinfo {volume} {88}},\ \bibinfo {pages}
  {905880408} (\bibinfo {year} {2022})}\BibitemShut {NoStop}%
\bibitem [{\citenamefont {Ito}\ and\ \citenamefont {Cappelli}(2009)}]{ito2009}%
  \BibitemOpen
  \bibfield  {author} {\bibinfo {author} {\bibfnamefont {T.}~\bibnamefont
  {Ito}}\ and\ \bibinfo {author} {\bibfnamefont {M.~A.}\ \bibnamefont
  {Cappelli}},\ }\href@noop {} {\bibfield  {journal} {\bibinfo  {journal}
  {Appl. Phys. Lett.}\ }\textbf {\bibinfo {volume} {94}},\ \bibinfo
  {pages} {211501} (\bibinfo {year} {2009})}\BibitemShut {NoStop}%
\bibitem [{\citenamefont {Marcovati}, \citenamefont {Ito},\ and\ \citenamefont
  {Cappelli}(2020)}]{marcovati2020}%
  \BibitemOpen
  \bibfield  {author} {\bibinfo {author} {\bibfnamefont {A.}~\bibnamefont
  {Marcovati}}, \bibinfo {author} {\bibfnamefont {T.}~\bibnamefont {Ito}}, \
  and\ \bibinfo {author} {\bibfnamefont {M.~A.}\ \bibnamefont {Cappelli}},\
  }\href@noop {} {\bibfield  {journal} {\bibinfo  {journal} {J. Appl. Phys.}\
  }\textbf {\bibinfo {volume} {127}},\ \bibinfo {pages} {223301} (\bibinfo
  {year} {2020})}\BibitemShut {NoStop}%
\bibitem [{\citenamefont {Marcovati}\ and\ \citenamefont
  {Cappelli}(2023)}]{marcovati2023}%
  \BibitemOpen
  \bibfield  {author} {\bibinfo {author} {\bibfnamefont {A.}~\bibnamefont
  {Marcovati}}\ and\ \bibinfo {author} {\bibfnamefont {M.~A.}\ \bibnamefont
  {Cappelli}},\ }\href@noop {} {\bibfield  {journal} {\bibinfo  {journal} {J.
  Appl. Phys.}\ }\textbf {\bibinfo {volume} {134}},\ \bibinfo {pages} {033306}
  (\bibinfo {year} {2023})}\BibitemShut {NoStop}%
\bibitem [{\citenamefont {Przybocki}\ and\ \citenamefont
  {Cappelli}(2023)}]{przybocki2023}%
  \BibitemOpen
  \bibfield  {author} {\bibinfo {author} {\bibfnamefont {R.~C.}\ \bibnamefont
  {Przybocki}}\ and\ \bibinfo {author} {\bibfnamefont {M.~A.}\ \bibnamefont
  {Cappelli}},\ }\href@noop {} {\bibfield  {journal} {\bibinfo  {journal}
  {Phys. Plasmas}\ }\textbf {\bibinfo {volume} {30}},\ \bibinfo {pages}
  {082101} (\bibinfo {year} {2023})}\BibitemShut {NoStop}%
%
\bibitem [{\citenamefont {Hecimovic}\ and\ \citenamefont
  {von Keudell}(2018)}]{hecimovic}%
  \BibitemOpen
  \bibfield  {author} {\bibinfo {author} {\bibfnamefont {A.}\ \bibnamefont
  {Hecimovic}}\ and\ \bibinfo {author} {\bibfnamefont {A.}\ \bibnamefont
  {von Keudell}},\ }\href@noop {} {\bibfield  {journal} {\bibinfo  {journal}
  {J. Phys. D: Appl. Phys.}\ }\textbf {\bibinfo {volume} {51}},\ \bibinfo
  {pages} {453001} (\bibinfo {year} {2018})}\BibitemShut {NoStop}%
%
%
\bibitem [{\citenamefont {Boeuf}(2023)}]{boeuf2023}%
  \BibitemOpen
  \bibfield  {author} {\bibinfo {author} {\bibfnamefont {J.~P.}~\bibnamefont
  {Boeuf}},\ }\href@noop {} {\bibfield  {journal} {\bibinfo  {journal}
  {Phys. Plasmas}\ }\textbf {\bibinfo {volume} {30}},\ \bibinfo {pages} {022112}
  (\bibinfo {year} {2023})}\BibitemShut {NoStop}%
%
%
\bibitem [{\citenamefont {Panjan}\ \emph {et~al.}(2015)\citenamefont
  {Panjan}, \citenamefont {Loquai}, \citenamefont {Klemberg-Saphieha},\ and\
  \citenamefont {Martinu}}]{panjan2015}%
  \BibitemOpen
  \bibfield  {author} {\bibinfo {author} {\bibfnamefont {M.}~\bibnamefont
  {Panjan}}, \bibinfo {author} {\bibfnamefont {S.}~\bibnamefont
  {Loquai}}, \bibinfo {author} {\bibfnamefont {J.~E.}~\bibnamefont {Klemberg-Sapieha}},
  \ and\ \bibinfo {author} {\bibfnamefont {L.}~\bibnamefont {Martinu}},\
  }\href@noop {} {\bibfield  {journal} {\bibinfo  {journal} {Plasma Sources Sci. Technol.}\
  }\textbf {\bibinfo {volume} {24}},\ \bibinfo {pages} {065010} (\bibinfo
  {year} {2015})}\BibitemShut {NoStop}%
%
%
\bibitem [{\citenamefont {Revel}\ \emph {et~al.}(2016)\citenamefont
  {Revel}, \citenamefont {Minea}, \ and\ \citenamefont {Tsikata}}]{revel2016}%
  \BibitemOpen
  \bibfield  {author} {\bibinfo {author} {\bibfnamefont {A.}~\bibnamefont
  {Revel}}, \bibinfo {author} {\bibfnamefont {T.}~\bibnamefont
  {Minea}}, \bibinfo {author}
  \ and\ \bibinfo {author} {\bibfnamefont {S.}~\bibnamefont {Tsikata}},\
  }\href@noop {} {\bibfield  {journal} {\bibinfo  {journal} {Phys. Plasmas}\
  }\textbf {\bibinfo {volume} {23}},\ \bibinfo {pages} {100701} (\bibinfo
  {year} {2016})}\BibitemShut {NoStop}%
%
\bibitem [{\citenamefont {Baltzis}\ (2008)\citenamefont
  {Baltzis}}]{baltzis2008}%
  \BibitemOpen
  \bibfield  {author} {\bibinfo {author} {\bibfnamefont {E.}~\bibnamefont
  {Baltzis}},\
  }\href@noop {} {\bibfield  {journal} {\bibinfo  {journal} {J. Eng. Sci. Technol. Rev.}\
  }\textbf {\bibinfo {volume} {1}},\ \bibinfo {pages} {83} (\bibinfo
  {year} {2008})}\BibitemShut {NoStop}%
%
\bibitem [{\citenamefont {Koshkarov}\ \emph {et~al.}(2019)\citenamefont
  {Koshkarov}, \citenamefont {Smolyakov}, \citenamefont {Raitses},\ and\
  \citenamefont {Kaganovich}}]{smolyakov_prl}%
  \BibitemOpen
  \bibfield  {author} {\bibinfo {author} {\bibfnamefont {O.}~\bibnamefont
  {Koshkarov}}, \bibinfo {author} {\bibfnamefont {A.}~\bibnamefont
  {Smolyakov}}, \bibinfo {author} {\bibfnamefont {Y.}~\bibnamefont {Raitses}},
  \ and\ \bibinfo {author} {\bibfnamefont {I.}~\bibnamefont {Kaganovich}},\
  }\href@noop {} {\bibfield  {journal} {\bibinfo  {journal} {Phys. Rev. Lett.}\
  }\textbf {\bibinfo {volume} {122}},\ \bibinfo {pages} {185001} (\bibinfo
  {year} {2019})}\BibitemShut {NoStop}%
%
\bibitem [{\citenamefont {Tyushev}\ \emph {et~al.}(2023)\citenamefont
  {Tyushev}, \citenamefont {Papahn Zadeh}, \citenamefont {Sharma}, \citenamefont {Sengupta}, \citenamefont {Raitses}, \citenamefont {Boeuf},\ and\
  \citenamefont {Smolyakov}}]{tyushev}%
  \BibitemOpen
  \bibfield  {author} {\bibinfo {author} {\bibfnamefont {M.}~\bibnamefont
  {Tyushev}}, \bibinfo {author} {\bibfnamefont {M.}~\bibnamefont
  {Papahn Zadeh}}, \bibinfo {author} {\bibfnamefont {V.}~\bibnamefont {Sharma}}, \bibinfo {author} {\bibfnamefont {M.}~\bibnamefont {Sengupta}}, \bibinfo {author} {\bibfnamefont {Y.}~\bibnamefont {Raitses}}, \bibinfo {author} {\bibfnamefont {J.~P.}~\bibnamefont {Boeuf}},
  \ and\ \bibinfo {author} {\bibfnamefont {A.}~\bibnamefont {Smolyakov}},\
  }\href@noop {} {\bibfield  {journal} {\bibinfo  {journal} {Phys. Plasmas}\
  }\textbf {\bibinfo {volume} {30}},\ \bibinfo {pages} {033506} (\bibinfo
  {year} {2023})}\BibitemShut {NoStop}%
%
\bibitem [{\citenamefont {Smolyakov}\ \emph {et~al.}(2017)\citenamefont
  {Smolyakov}, \citenamefont {Chapurin}, \citenamefont {Frias}, \citenamefont
  {Koshkarov}, \citenamefont {Romadanov}, \citenamefont {Tang}, \citenamefont
  {Umansky}, \citenamefont {Raitses}, \citenamefont {Kaganovich},\ and\
  \citenamefont {Lakhin}}]{smolyakov2017}%
  \BibitemOpen
  \bibfield  {author} {\bibinfo {author} {\bibfnamefont {A.}\ \bibnamefont
  {Smolyakov}}, \bibinfo {author} {\bibfnamefont {O.}~\bibnamefont {Chapurin}},
  \bibinfo {author} {\bibfnamefont {W.}~\bibnamefont {Frias}}, \bibinfo
  {author} {\bibfnamefont {O.}~\bibnamefont {Koshkarov}}, \bibinfo {author}
  {\bibfnamefont {I.}~\bibnamefont {Romadanov}}, \bibinfo {author}
  {\bibfnamefont {T.}~\bibnamefont {Tang}}, \bibinfo {author} {\bibfnamefont
  {M.}~\bibnamefont {Umansky}}, \bibinfo {author} {\bibfnamefont
  {Y.}~\bibnamefont {Raitses}}, \bibinfo {author} {\bibfnamefont
  {I.}~\bibnamefont {Kaganovich}}, \ and\ \bibinfo {author} {\bibfnamefont
  {V.}~\bibnamefont {Lakhin}},\ }\href@noop {} {\bibfield  {journal} {\bibinfo
  {journal} {Plasma Phys. Control. Fusion}\ }\textbf {\bibinfo {volume} {59}},\
  \bibinfo {pages} {014041} (\bibinfo {year} {2017})}\BibitemShut {NoStop}%
\bibitem [{\citenamefont {Mikellides}\ and\ \citenamefont
  {Ortega}(2020)}]{mikellides}%
  \BibitemOpen
  \bibfield  {author} {\bibinfo {author} {\bibfnamefont {I.}~\bibnamefont
  {Mikellides}} \ and\  \bibinfo {author} {\bibfnamefont {A.}~\bibnamefont {Ortega}},\ }\href@noop {}
  {\bibfield  {journal} {\bibinfo  {journal} {Phys. Plasmas}\
  }\textbf {\bibinfo {volume} {27}},\ \bibinfo {pages} {100701} (\bibinfo {year}
  {2020})}\BibitemShut {NoStop}%
\bibitem [{\citenamefont {Norgren}\ \emph {et~al.}(2012)\citenamefont {Norgren},
  \citenamefont {Vaivads}, \citenamefont {Khotyaintsev},\ and\ \citenamefont
  {Andre}}]{norgren}%
  \BibitemOpen
  \bibfield  {author} {\bibinfo {author} {\bibfnamefont {C.}~\bibnamefont
  {Norgren}}, \bibinfo {author} {\bibfnamefont {A.}~\bibnamefont {Vaivads}},
  \bibinfo {author} {\bibfnamefont {Y.~V.}~\bibnamefont {Khotyaintsev}}, \ and\ \bibinfo
  {author} {\bibfnamefont {M.}~\bibnamefont {Andre}},\ }\href@noop {}
  {\bibfield  {journal} {\bibinfo  {journal} {Phys. Rev. Lett.}\
  }\textbf {\bibinfo {volume} {109}},\ \bibinfo {pages} {055001} (\bibinfo {year}
  {2012})}\BibitemShut {NoStop}%
\bibitem [{\citenamefont {Krall}\ and\ \citenamefont
  {Liewer}(1971)}]{krall}%
  \BibitemOpen
  \bibfield  {author} {\bibinfo {author} {\bibfnamefont {N.}~\bibnamefont
  {Krall}} \ and\  \bibinfo {author} {\bibfnamefont {P.}~\bibnamefont {Liewer}},\ }\href@noop {}
  {\bibfield  {journal} {\bibinfo  {journal} {Phys. Rev. A}\
  }\textbf {\bibinfo {volume} {4}},\ \bibinfo {pages} {2094} (\bibinfo {year}
  {1971})}\BibitemShut {NoStop}%
\bibitem [{\citenamefont {Huba}\ \emph {et~al.}(2021)\citenamefont {Huba},
  \citenamefont {Gladd}, and\ \citenamefont {Papadopoulos}}]{huba1978}%
  \BibitemOpen
  \bibfield  {author} {\bibinfo {author} {\bibfnamefont {J.~D.}~\bibnamefont
  {Huba}} \, \bibinfo {author} {\bibfnamefont {N.~T.}~\bibnamefont
  {Gladd}} \, and \bibinfo {author} {\bibfnamefont {K.}~\bibnamefont {Papadopoulos}},\ }\href@noop {}
  {\bibfield  {journal} {\bibinfo  {journal} {J. Geophys. Res.}\
  }\textbf {\bibinfo {volume} {83}},\ \bibinfo {pages} {5217} (\bibinfo {year}
  {1978})}\BibitemShut {NoStop}%
\bibitem [{\citenamefont {Davidson}\ \emph {et~al.}(1976)\citenamefont {Davidson},
  \citenamefont {Gladd}, \citenamefont {Wu}, and\ \citenamefont {Huba}}]{davidson1976}%
  \BibitemOpen
  \bibfield  {author} {\bibinfo {author} {\bibfnamefont {R.~C.}~\bibnamefont
  {Davidson}} \, \bibinfo {author} {\bibfnamefont {N.~T.}~\bibnamefont
  {Gladd}} \, \bibinfo {author} {\bibfnamefont {C.~S.}~\bibnamefont
  {Wu}} \, and \bibinfo {author} {\bibfnamefont {J.~D.}~\bibnamefont {Huba}},\ }\href@noop {}
  {\bibfield  {journal} {\bibinfo  {journal} {Phys. Rev. Lett.}\
  }\textbf {\bibinfo {volume} {37}},\ \bibinfo {pages} {750} (\bibinfo {year}
  {1976})}\BibitemShut {NoStop}%
\bibitem [{\citenamefont {Davidson}\ and\ \citenamefont
  {Gladd}(1971)}]{davidson1975}%
  \BibitemOpen
  \bibfield  {author} {\bibinfo {author} {\bibfnamefont {R.~C.}~\bibnamefont
  {Davidson}} \ and\  \bibinfo {author} {\bibfnamefont {N.~T.}~\bibnamefont {Gladd}},\ }\href@noop {}
  {\bibfield  {journal} {\bibinfo  {journal} {Phys. Fluids}\
  }\textbf {\bibinfo {volume} {18}},\ \bibinfo {pages} {1327} (\bibinfo {year}
  {1975})}\BibitemShut {NoStop}%
\bibitem [{\citenamefont {Kawashima}\ \emph {et~al.}(2018)\citenamefont     {Kawashima},
  \citenamefont {Hara}, \ and\ \citenamefont
  {Komurasaki}}]{hara}%
  \BibitemOpen
  \bibfield  {author} {\bibinfo {author} {\bibfnamefont {R.}~\bibnamefont
  {Kawashima}}, \bibinfo {author} {\bibfnamefont {K.}~\bibnamefont {Hara}}, and\ \bibinfo
  {author} {\bibfnamefont {K.}~\bibnamefont {Komurasaki}},\ }\href@noop {}
  {\bibfield  {journal} {\bibinfo  {journal} {Plasma Sources Sci. Technol.}\
  }\textbf {\bibinfo {volume} {27}},\ \bibinfo {pages} {035010} (\bibinfo {year}
  {2018})}\BibitemShut {NoStop}%
\bibitem [{\citenamefont {Fernandez}\ \emph {et~al.}(2008)\citenamefont{Fernandez},\citenamefont {Scharfe}, \citenamefont {Thomas}, \citenamefont {Gascon},\ and\ \citenamefont
  {Cappelli}}]{fernandez2008}%
  \BibitemOpen
  \bibfield  {author} {\bibinfo {author} {\bibfnamefont {E.}~\bibnamefont
  {Fernandez}}, \bibinfo {author} {\bibfnamefont {M. K.}~\bibnamefont {Scharfe}},
  \bibinfo {author} {\bibfnamefont {C. A.}~\bibnamefont {Thomas}},  \bibinfo {author} {\bibfnamefont {N.}~\bibnamefont {Gascon}}, \ and\ \bibinfo
  {author} {\bibfnamefont {M. A.}~\bibnamefont {Cappelli}},\ }\href@noop {}
  {\bibfield  {journal} {\bibinfo  {journal} {Phys. Plasmas}\
  }\textbf {\bibinfo {volume} {15}},\ \bibinfo {pages} {012102} (\bibinfo {year}
  {2008})}\BibitemShut {NoStop}%
\bibitem [{\citenamefont {Mitic}\ \emph {et~al.}(2021)\citenamefont {Mitic},
  \citenamefont {Moreno}, \citenamefont {Arnas},\ and\ \citenamefont
  {Couedel}}]{mitic2021}%
  \BibitemOpen
  \bibfield  {author} {\bibinfo {author} {\bibfnamefont {S.}~\bibnamefont
  {Mitic}}, \bibinfo {author} {\bibfnamefont {J.}~\bibnamefont {Moreno}},
  \bibinfo {author} {\bibfnamefont {C.}~\bibnamefont {Arnas}}, \ and\ \bibinfo
  {author} {\bibfnamefont {L.}~\bibnamefont {Couedel}},\ }\href@noop {}
  {\bibfield  {journal} {\bibinfo  {journal} {Eur. Phys. J. D}\
  }\textbf {\bibinfo {volume} {75}},\ \bibinfo {pages} {240} (\bibinfo {year}
  {2021})}\BibitemShut {NoStop}%
\end{thebibliography}
\end{document}